\documentclass[12pt,thmsa]{article}
\usepackage{amssymb}
\usepackage{sw20aip}

\input tcilatex
\QQQ{Language}{
American English
}

\begin{document}

\topmargin 0pt \oddsidemargin 5mm

\setcounter{page}{1}

\hspace{8cm}Preprint YerPhI-1521(21)-98

\begin{quotation}
\hspace{8cm}{} \vspace{2cm}
\end{quotation}

\begin{center}
\textbf{An Explicit Solution for Nonlinear Plasma Waves of Arbitrary
Amplitude}\\\vspace {5mm}

A. G. Khachatryan and S. S. Elbakian\\\vspace{1cm} \emph{Yerevan Physics
Institute, Alikhanian Brothers St. 2, Yerevan 375036, Republic of Armenia}\\%
E-mail: khachatr@moon.yerphi.am
\end{center}

\vspace {5mm} \centerline{{\bf{Abstract}}}

\begin{quotation}
Based on the known implicit solution for nonlinear plasma waves, an explicit
solution was obtained in the form of decomposition into harmonics. The
solution obtained exhibits a mechanism for steepening of nonlinear plasma
wave as a result of increasing contribution of harmonics and may be used in
theoretical studies of processes involving nonlinear plasma waves.

\newpage\ 
\end{quotation}

In the present work an explicit solution for one-dimensional nonlinear
plasma waves was obtained. Below, prior to the discussion of the solution 
\textit{per se, }we shall give a formal mathematical description of the
method for obtaining explicit solutions in case when the implicit solutions
are known.

\textbf{1)} In many nonlinear problems of physics the solutions are sought
in expressions of the following form

\begin{equation}
x=f(y),  \tag{1}
\end{equation}
where $f(y)$ is some nonlinear function. The variable $x$ may be e.g., the
coordinate or the time, while as $y$ the electric field or magnetic field
strengths, the velocity of the continuum etc. may serve. The expressions of
the type (1) are usually termed as implicitly given with respect to the
quantity $y$. This equation may describe, for instance, some nonlinear
oscillation phenomena, the results of experimental observations, the
distribution of electromagnetic field in a cavity or the time dependence of
electric field strength in the receiving aerial. The implicit solutions of
the type (1) are usually highly inconvenient and the knowledge of an
explicit solution of the type $y=y\left( x\right) $ is much preferable. It
is frequently required to have an explicit solution in the form of
decomposition into harmonics or of a spectrum (e.g., as a series or the
Fourier integral), a series expansion in some set of functions etc. The
method that is described below will permit the reconstruction from the
relation (1) of explicit representations for some function $g(y)$ (in
particular $g=y$) in the form of series expansion or an integral.

Let some function $q(x)$ be given. The transform of this function is called
the quantity (see, e.g., [1]) 
\begin{equation}
Q(\alpha )=L_x[K_{+}(\alpha ,x)q(x)],  \tag{2}
\end{equation}
where $L_x$ is an operator that usually stands for the integration over the $%
x$ variable, and the function $K_{+}(\alpha ,x)$ is termed as the kernel of
transformation. If $Q(\alpha )$ is known, then one can determine the
function $q(x)$ by means of an inverse transformation

\begin{equation}
q(x)=L_\alpha [K_{-}(\alpha ,x)Q(\alpha )].  \tag{3}
\end{equation}
In (3) $L_\alpha $ is the operator of inverse transformation implying the
summation or integration over the parameter $\alpha ,$ $K_{-}(\alpha ,x)$ is
the kernel of inverse transformation. The expressions (2) and (3) describe
the Fourier, Hankel, Laplace, Mellin transforms etc. Suppose that (1) is
given and it is required to find an explicit expression for some given
function $g(y),$ i.e., to find $g=g(x).$ We have from (1):

\begin{equation}
L_\alpha [K_{-}(\alpha ,x)R(\alpha )]=L_\alpha [K_{-}(\alpha ,f)R(\alpha )] 
\tag{4}
\end{equation}
Here $R(\alpha )$ is not known. The right hand side of the expression (4) is
a transform of the type (3) for some function of $y.$ We shall require that
this function be the one of interest to us, $g(y).$ We can then determine
the unknown function $R(\alpha ).$ Thus,

\begin{center}
\begin{equation}
g(y)=L_\alpha [K_{-}(\alpha ,f)R(\alpha )],  \tag{5}
\end{equation}
\end{center}

\begin{equation}
R(\alpha )=L_f[K_{+}(\alpha ,f(y))g(y)].  \tag{6}
\end{equation}
Since the left hand side of (4) is also equal to $g(y)$, then taking into
account (6) we obtain the required explicit representation for $g(y)$ as a
function of $x$:

\begin{equation}
g(x)=L_\alpha [K_{-}(\alpha ,x)L_f[K_{+}(\alpha ,f(y))g(y)].  \tag{7}
\end{equation}
So, if we have the implicit expression of the form (1), we can find the
explicit representation $g\left( x\right) $ for some function $g(y)$ by
means of the formula (7). Although the method under consideration in this or
that form is known in mathematics, e.g., in the theory of Bessel functions
[2], however, it is rarely applied in physics. The approach similar to the
one discussed above may be of help also at calculations of intricate sums
over Bessel functions [3].

\textbf{2) }Now, we shall determine the explicit solution for the nonlinear
one-dimensional plasma waves in cold plasma. These waves are described by a
set of equations consisting of the equation of motion and continuity
equation for plasma electrons as well as of the Poisson equation:

\begin{equation}
\frac{\partial v_e}{\partial t}+v_e\frac{\partial v_e}{\partial Z}=\frac{%
\left| e\right| }{m_e}\cdot \frac{\partial \varphi }{\partial Z},  \tag{8}
\end{equation}

\begin{center}
\begin{equation}
\frac{\partial n_e}{\partial t}+\frac \partial {\partial Z}(n_ev_e)=0, 
\tag{9}
\end{equation}
\end{center}

\begin{equation}
\frac{\partial ^2\varphi }{\partial Z^2}=-4\pi \left| e\right| (n_o-n_e). 
\tag{10}
\end{equation}
where $v_e$ and $n_e$ are the velocity and density of electrons, $n_o$ is
the equilibrium value of density, $\varphi $ - the electric potential that
is related with the electric field strength by means of the formula $%
E=-\partial \varphi \diagup \partial Z,$ $e$ and $m_e$ being the electron
charge and mass respectively. For steady nonlinear waves propagating with
the phase velocity $v_{ph},$ one can obtain from (8)-(10) the following
equation for the potential [4]:

\begin{equation}
\frac{\omega _{pe}^2}{v_{ph}^2}\cdot \frac{d^2\varphi }{d\tau ^2}+4\pi
n_o\left| e\right| \left[ 1-\frac{v_{ph}}{(2\left| e\right| \varphi \diagup
m_e)^{1/2}}\right] =0,  \tag{11}
\end{equation}
where $\tau =\omega _{pe}(t-Z\diagup v_{ph}),$ $\omega _{pe}=(4\pi
n_oe^2\diagup m_e)^{1/2}$ is the electron plasma frequency. Multiplying (11)
by $d\varphi /d\tau $ one can find the first integral in the form $(d\varphi
/d\tau )^2\diagup 2+U(\varphi )=const.$ The integration of the latter
expression gives an implicit solution of the problem [4] (see also [5]):

\begin{equation}
\tau =f(y)=-\arccos y-A(1-y^2)^{1/2}.  \tag{12}
\end{equation}
In (12) $-1\leqslant y=(\Phi ^{1/2}-1)\diagup A\leqslant 1,$ $%
(1-A)^2\leqslant \Phi =2\left| e\right| \varphi \diagup
(m_ev_{ph}^2)\leqslant $ $(1+A)^2,$ $\Phi $ is the dimensionless electric
potential , $A=v_m/v_{ph}\leqslant 1_{,}$ where $v_m$ is the maximum
velocity of electrons in the wave. We try to obtain the explicit expression
for the dimensionless potential $\Phi (\tau )$ as an series expansion into
harmonics , i.e., in the form of Fourier series (note that the procedure to
be described below may be pursued also for another quantity, e.g., for $\Phi
^{1/2}$). Since according to (12) $\Phi (\tau )$ is an even function, we
shall seek the explicit expression in the form of Fourier series that
comprises the cosines only. Then, pursuant to the above method, we have from
(12),

\begin{equation}
\stackrel{\infty }{\stackunder{n=0}{\sum }}a_n\cos (n\tau )=\stackunder{n=0}{%
\stackrel{\infty }{\sum }}a_n\cos (nf)=\Phi .  \tag{13}
\end{equation}
Hence, according to the Fourier transformation

\begin{equation}
a_n=(2/\pi )\stackunder{0}{\stackrel{\pi }{\int }}\Phi \cos (nf)df=  \tag{14}
\end{equation}
\[
(2/\pi )\stackunder{-1}{\stackrel{1}{\int }}(1+Ay)^3(1-y^2)^{1/2}\cos
\left\{ n[\arccos y+A(1-y^2)^{1/2}]\right\} dy
\]
Making the substitution $y=\cos \Psi $ we have from (14)

\begin{equation}
a_n=-(2/\pi )\stackunder{0}{\stackrel{\pi }{\int }}(1+A\cos \Psi )^3\cos
[n(\Psi +A\sin \Psi )]d\Psi .  \tag{15}
\end{equation}
Using the well known relations for Bessel functions of integer order (see,
e.g., [1])

\[
J_n(-x)=(-1)^nJ_n(x), 
\]

\[
J_n(x)=(1/\pi )\stackunder{0}{\stackrel{\pi }{\int }}\cos (n\Psi -x\sin \Psi
)d\Psi , 
\]
as well as the known recurrent expressions for the Bessel and
trigonometrical functions, we obtain from (15)

\begin{equation}
a_n=-(-1)^n(4/n^2)J_n(nA).  \tag{16}
\end{equation}
Then, according to (13), the required expansion into harmonics for the
dimensionless potential will take the form:

\begin{equation}
\Phi (\tau )=-4\stackunder{n=0}{\stackrel{\infty }{\sum }}(-1)^n\frac{J_n(nA)%
}{n^2}\cos (n\tau ).  \tag{17}
\end{equation}
And for the strength of the electric field we have an explicit expression

\begin{center}
\begin{equation}
E(\tau )=-\frac{\partial \varphi }{\partial Z}=\frac{m_e\omega _{pe}v_{ph}}{%
2\left| e\right| }\frac{d\Phi }{d\tau }=\frac{m_e\omega _{pe}v_{ph}}{\left|
e\right| }\cdot 2\stackunder{n=1}{\stackrel{\infty }{\sum }}(-1)^n\frac{%
J_n(nA)}n\sin (n\tau ).  \tag{18}
\end{equation}
\end{center}

The nonlinear plasma waves excited by bunches of charged particles were
formerly studied by simulation of the initial equation (8) - (10) [6]. The
plots for the potential and strength of electric field in the plasma wave
obtained by (17) and (18), and those obtained by means of computer
simulation of nonlinear plasma wave excitation [6] practically coincided in
case when the amplitudes of waves were equal . For weak plasma waves, when $%
A=v_m/v_{ph}\ll 1,$ the series expansions in (17) and (18) are rapidly
converging. Hence, in the linear case we arrive at the well known result: $%
\varphi \approx \varphi _{\max }\cos \tau ;$ $E\approx E_{\max }\sin \tau .$
The contribution of the harmonics increases with the amplitude and leads to
characteristic steepening of the wave and eventually to the wavebreaking at $%
A=1,$ $E_{\max }=m_e\omega _{pe}v_{ph}/|e|.$

Thus, the nonlinear variation of the shape of plasma wave is connected with
the increase in contribution of harmonics. In the theory of processes
involving weakly nonlinear plasma waves one can confine to the consideration
of a few (say, the first two or three) terms in the series expansions (17)
and (18).

This work was in part supported by the International Science and Technology
Center, Project A-013.

\begin{center}
\textbf{REFERENCES}
\end{center}

[1] G. A. Korn and T. M. Korn, \textit{Mathematical Handbook} (McGraw-Hill,
New York, 1968), chap. 10.

[2] G. N. Watson, \textit{A treatise on the theory of Bessel functions},
(1945), chap. 17.

[3] A. G. Khachatryan, A. Ts. Amatuni, S. S. Elbakian, E. V. Sekhpossian,
Plasma Phys. Rep. \textbf{22}, 576 (1996).

[4] A. I. Akhiezer and G. Ya. Ljubarski, Doklady AN SSSR \textbf{80}, 193
(1951).

[5] S. F. Smerd, Nature \textbf{175}, 297 (1955).

[6] A. G. Khachatryan, Phys. Plasmas \textbf{4}, 4136 (1997).

\end{document}